\documentclass[prl,aps,twocolumn,reprint]{revtex4-1}

\usepackage{graphicx} 
\usepackage[usenames]{color}
\usepackage{amsmath,amssymb}
\usepackage{gensymb}
\usepackage{natbib}
\begin{document} 

\title{Formation of surface nanobubbles and universality of their contact angles: A molecular dynamics approach}
\author{Joost H. Weijs}
\affiliation{Physics of Fluids Group and J. M. Burgers Centre for Fluid Dynamics,
University of Twente, P.O. Box 217, 7500 AE Enschede, The Netherlands}
\author{Jacco H. Snoeijer}
\affiliation{Physics of Fluids Group and J. M. Burgers Centre for Fluid Dynamics,
University of Twente, P.O. Box 217, 7500 AE Enschede, The Netherlands}
\author{Detlef Lohse}
\email{d.lohse@utwente.nl}
\affiliation{Physics of Fluids Group and J. M. Burgers Centre for Fluid Dynamics,
University of Twente, P.O. Box 217, 7500 AE Enschede, The Netherlands}

\date{\today} 

\begin{abstract}
We study surface nanobubbles using molecular dynamics simulation of ternary (gas, liquid, solid) systems of Lennard-Jones fluids. They 
form for sufficiently low gas solubility in the liquid, i.e.,
for large relative gas concentration.
For strong enough gas-solid attraction, the surface nanobubble is sitting on a gas layer, 
which forms in between the liquid and the solid. This gas layer is the reason for the universality of the 
contact angle, which we calculate from the microscopic parameters.
Under the present equilibrium conditions the nanobubbles dissolve within less of a microsecond,
consistent with the view  that the experimentally found nanobubbles are stabilized by a nonequilibrium mechanism.
\end{abstract} 

\maketitle

When liquid comes into contact with a solid, nanoscopic gaseous bubbles can  form at the interface: surface nanobubbles~\cite{Hampton,Seddonrev11,Craig11}. 
These bubbles were discovered about 15 years ago, after Parker~\emph{et al.} predicted their existence to explain the long-ranged attraction between hydrophobic surfaces in water~\cite{Parker94}. 
Many Atomic Force Microscopy (AFM) 
and spectroscopy measurements have since then confirmed the existence of spherical cap-shaped, gaseous bubbles at the liquid-solid interface. 

Various open questions remain about surface nanobubbles, and in this Letter we will address three crucial ones:
(i) How do surface nanobubbles form?
This question is difficult to answer by experimental means, since the formation process is too fast to be observed by AFM.
(ii) A second question regards the contact angle of surface nanobubbles which is found to disagree with Young's law: all recorded nanobubbles have a much lower gas-side contact angle than expected, and seem to be universal within 20 degrees.
(iii) Finally, AFM showed that surface nanobubbles can be stable for hours or even days, whereas the pressure inside these bubbles due to their small radius of curvature ($R_c \sim 100$~nm) would be several atmospheres due to the Laplace pressure: $\Delta p=2\gamma /R_c$, with $\gamma$ the liquid-vapour surface tension. 
A simple calculation then shows that surface nanobubbles should dissolve within microseconds, which is 9 to 10 orders of magnitudes off with respect to the experimental data.

\begin{figure}[t]
\begin{center}
  \includegraphics[width=70mm]{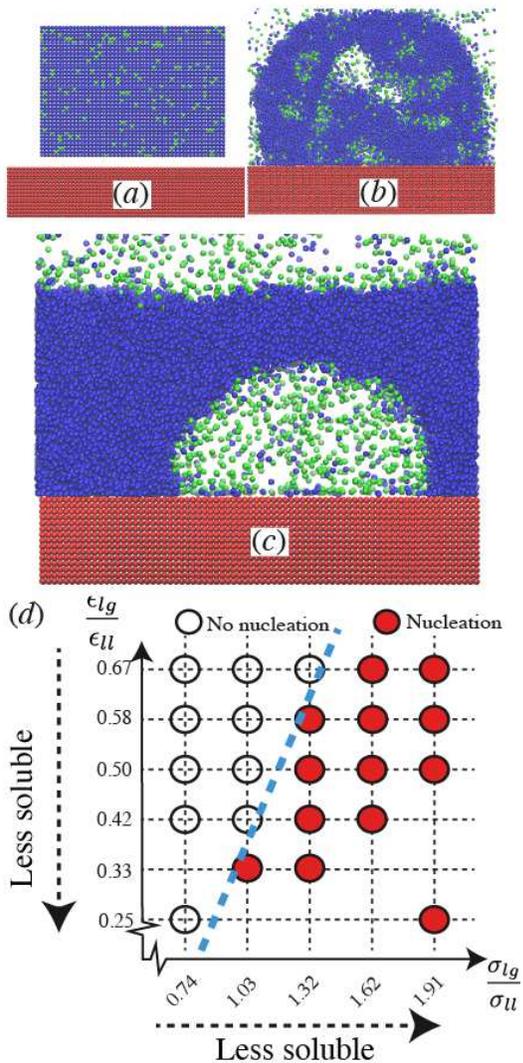}
  \caption{\label{fig:phasespace}(Color) a) Initial conditions: a liquid layer (blue) is placed on top of a solid substrate (bottom, red). Gas is dissolved inside the liquid layer (green). 
  b) Simulation ($\epsilon_{lg}/\epsilon_{ll}=0.58$, $\sigma_{lg}/\sigma_{ll}=1.32$) after about 0.1 ns: nucleation occurs. 
  c) $t=10$~ns: a surface nanobubble has formed.
  d) Parameter space where the solubility of the gas was tuned through the parameters $\epsilon_{lg}/\epsilon_{ll}$ and $\sigma_{lg}/\sigma_{ll}$. As the gas becomes increasingly soluble (going up, left in Fig.~\ref{fig:phasespace}c) a sudden transition takes place where no nanobubbles nucleate. The gas then remains in a dissolved state and partially escapes to the gas-phase above the liquid layer until equilibrium is reached.}
\end{center}
\end{figure}

In this paper, we use Molecular Dynamics simulations (MD) to study surface nanobubbles in simple fluids. Using MD, we are able to answer questions (i) and (ii), and provide important information with respect to question (iii). 
MD are well-suited for nanobubbles, because the temporal resolution is of order fs, and since all atom's motions are resolved, the spatial resolution is intrinsically high enough to resolve nanobubbles. This atomistic model allows to study microscopic details that are inaccessible by experimental means and standard continuum mechanics. 
Figure~\ref{fig:phasespace} shows how surface nanobubbles form in a typical simulation of a liquid containing gas. The gas will homogeneously nucleate to form a bubble, which subsequently attaches to the wall. We will analyze the nucleation process in detail and quantify  how the contact angle of the bubble changes upon varying gas solubility. 
The enhanced gas concentration (``gas-enrichment layer'') at the solid-liquid interface, which is strongest at hydrophobic substrates, will turn out to play a key role to account for the universality of the contact angle.

\paragraph{Numerical setup.} 
The studies in this paper are performed using simple fluids, which contain no molecules but rather separate atoms that interact with each other through the Lennard-Jones (LJ12-6) potential:
\begin{equation}
U=4\epsilon_{ij}\left[ \left(\frac{\sigma_{ij}}{r}\right)^{12}  - \left(\frac{\sigma_{ij}}{r}\right)^6  \right]\;.
\end{equation}
Here, $\epsilon_{ij}$ and $\sigma_{ij}$ are the interaction strength and range between particles $i$~and~$j$, respectively.
All simulations were performed using the \textsc{Gromacs}-software package and were done at constant temperature, volume and number of particles ($T$, $V$, and $N$ constant). 
The augmented Berendsen thermostat described in ref.~\onlinecite{Bussi07} was used in all simulations.
We verified that this thermostat yields the same result as the Nos{\'e}-Hoover thermostat~\footnote{See supplementary material}.
In all simulations three types of particles were used: fluid I, fluid II, and solid particles. 
The fluid particles (I~\&~II) behave like ordinary particles in a MD-simulation and thus move around the system. 
Contrarily, the solid particles are constrained to their initial position throughout time and constitute the immobile substrate.
The interaction parameters of the fluids are chosen such that at the  temperature considered ($T=300$~K) fluid I is in the liquid state and fluid II in the gas state, and they will be referred to by these states throughout the rest of the paper.
These interaction parameters are: $(\sigma_{ss},\sigma_{ll},\sigma_{gg}) = (0.34,0.34,0.50)$~nm, $(\epsilon_{ss},\epsilon_{ll},\epsilon_{gg})=(1.2,1.2,0.4)k_BT$, with $k_B$ Boltzmann's constant.
For cross-interactions we use: $\sigma_{ij}=(\sigma_{ii}+\sigma_{jj})/2$, and $(\epsilon_{sl},\epsilon_{sg},\epsilon_{lg})=(0.8,0.8,0.7)k_BT$, unless otherwise stated in the text. 
The cut-off length of the potential function was set at $r_c=5\sigma=1.7$nm.

The time step for the simulations is $dt=\tau/400\approx 2$~fs, where $\tau$ is a characteristic timescale of atomic motion $\tau = \sigma_{ll}\sqrt{m/\epsilon_{ll}}$, with $m$ the mass of the liquid particles (20 a.m.u.).
The initial conditions are shown in Fig.~\ref{fig:phasespace}a: on top of the substrate we place a layer of liquid with dissolved gas.
Periodic boundary conditions are present in all directions ($x$, $y$, $z$); the resulting nanobubbles are approximately 20-30~nm wide, and 10-20 nm high, depending on the contact angle.
The formation and behaviour of bubbles was found to be independent of simulation box-size, which was set at 40x40x5.5~nm$^3$.

\paragraph{Bubble nucleation. } What determines whether nanobubbles form?
We address this question by varying the relative interaction strength and the relative interaction size.
 We then  explore the parameter space to see under what conditions  nanobubbles form. The result is shown 
 in Fig.~\ref{fig:phasespace}d.
Decreasing $\epsilon_{lg}/\epsilon_{ll}$ (going down in Fig.~\ref{fig:phasespace}d) results in a lower solubility of gas in the liquid, and since the absolute concentration of gas is kept constant this means that the liquid becomes more and more supersaturated.
Eventually, homogeneous nucleation occurs and  
a nanobubble forms 
in the bulk liquid phase, which finally attaches to the surface.
Increasing $\sigma_{lg}/\sigma_{ll}$ (going right in Fig.~\ref{fig:phasespace}d) leads to the same effect: due to the increased size of the gas atoms it becomes energetically less favourable to remain dissolved in the liquid phase.
Eventually, when the gas molecules are large enough nanobubbles form due to the supersaturation of gas in the liquid.
From this, we can conclude that a local supersaturation of gas inside the liquid is a possible mechanism to generate surface nanobubbles. 
These results are consistent with the experimental findings in ref.~\onlinecite{Seddon11}, where it was
reported that nanobubble formation strongly depends on the (relative) gas concentration in the liquid.
Although the concentration required to spontaneously form nanobubbles is far greater than the saturation concentration, we point out that during deposition of liquid on a substrate gas can be trapped leading to very high local
 transient concentrations which would not be reflected in measurements of the global gas concentration in the liquid.
In fact, numerous experimental papers have pointed out that the method of deposition is of great importance for achieving surface nanobubbles~\cite{Zhang04}. 
We have to note, however, that other mechanisms not considered here can also induce the formation of nanobubbles (e.g. heterogeneous nucleation, bulk desorption of gas from micropancakes \cite{Zhang07,Seddon11}). The nanobubbles produced in our simulations are found to be reproducible and, at the very least, can be studied regarding their shape and stability.

\begin{figure}[t]
\begin{center}
  \includegraphics[width=82mm]{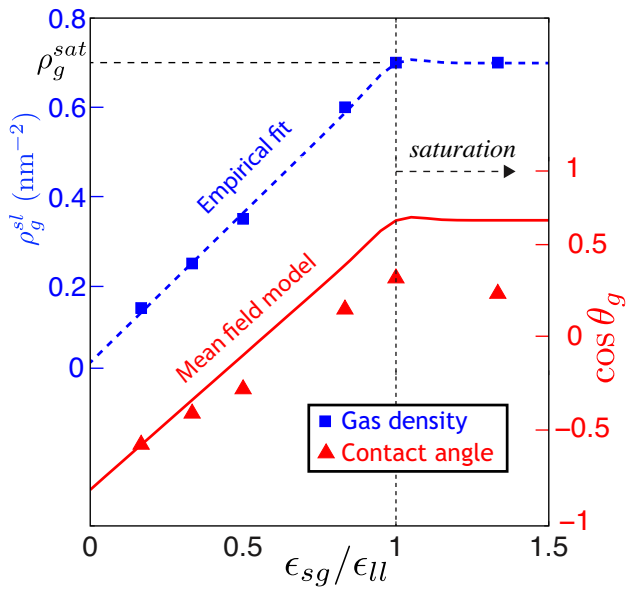}\hspace{3.5cm}
  \caption{\label{fig:epsgas}(Color) The effect of an enhanced gas-solid interaction strength. As the interaction strength $\epsilon_{sg}/\epsilon_{ll}$ is increased, the adsorbed gas density ($\rho_g^{sl}$, blue squares) increases as well until a saturation limit. As the adsorbed gas density increases, the gas-side contact angle $\theta$ lowers ($\cos \theta$ indicated by red triangles). The red solid line is a fit to the mean-field expression 
  (\ref{eq:theta}) taking into account the screening of the adsorbed gas (see text). 
    }
\end{center}
\end{figure}


\paragraph{Universal contact angle.} 
Now that we can simulate nanobubbles, we focus on the universal contact angle of surface nanobubbles found in experiments~\cite{Ishida00,Yang03,Zhang06ca}.
{For this we use similar initial conditions as in Fig.~\ref{fig:phasespace}a, with a region in the liquid with a very high gas concentration, providing us with control over where the nanobubbles form, and how much gas they contain (about 1,000 atoms).} 
We measure the gas-side contact angle{\footnote{Similarly as done in ref.~\cite{Weijs10} for nanodroplets.}}
 at varying values of the solid-gas interactions $\epsilon_{sg}$.
As can be seen in Fig.~\ref{fig:epsgas} (triangles), we observe that the gas-side contact angle of the nanobubble decreases (i.e. the nanobubble becomes flatter), as the solid-gas interaction is increased. This trend saturates around $\epsilon_{sg}/\epsilon_{ll} \approx 1$, where the gas-solid attraction matched the liquid-liquid attraction \footnote{A realistic value for the energy between gaseous argon and individual carbon atoms in graphene is: $\epsilon_{Ar\;-\;C} = 0.2$~$k_bT$. For the energies between the oxygens of two water molecules: $\epsilon_{O_l\;-\;O_l} = 0.26$~$k_bT$.\cite{GROMOSff}}. 
The observed saturation contact angle (75$^\circ$) is close to the contact angle of nanobubbles that is found in experiments (60$^\circ$).
On the same figure we show the evolution of the 2D number density of gas concentration at the wall inside the liquid phase, $\rho_g^{sl}$. Remarkably, this concentration exhibits a trend that is very similar to that of the contact angle. Stronger solid-gas interactions lead to a planar area of high gas concentration at the solid-liquid interface, which is called a gas-enrichment layer and which has been observed to exist in both simple liquids as well as real liquids~\cite{Dammer,Bratko08,Sendner09}. In experiments high-density gas adsorbates (micropancakes) have also been observed~\cite{Zhang07,Seddon11}.
Figure~\ref{fig:epsgas} shows that the adsorbate density increases with $\epsilon_{sg}$, until it finally saturates to a 2D number density of $\rho_g^{sl}=0.7$ atoms per nm$^2$ in the first gas layer above the substrate. 

The increase of the gas density near the wall is indeed the origin of the flattening of the nanobubbles. Namely, the presence of the gas weakens the attractive interaction between solid and liquid molecules: the liquid does not `feel' a solid half-space anymore, but there is now a dense gas-layer that effectively renders the wall more hydrophobic. This effect can be quantified using the approximate equation for the contact angle~\cite{Rowlinson/Widom,BauerDietrich,Snoeijer}, 
\begin{equation}\label{eq:theta}
\cos\theta_g=1-2\frac{\rho_s \epsilon_{sl}}{\rho_l \epsilon_{ll}},
\end{equation} 
which can be 
obtained from a mean field argument. This expression contains only the solid and liquid densities $\rho_s$, $\rho_l$, and the solid-liquid and liquid-liquid interactions $\epsilon_{sl}$, $\epsilon_{ll}$. The vapour phase has a negligible contribution ($\rho_v$ is small compared to $\rho_l$ and $\rho_s$) and the solid-solid interaction is irrelevant since the solid is non-deformable. In the case of a dense adsorbate, the attraction $\epsilon_{sl}$ is reduced to an effective interaction $\tilde{\epsilon}_{sl}$, with $\tilde{\epsilon}_{sl} < \epsilon_{sl}$. In addition, the adsorbate density is lower than the original solid density, due to the large size of the gas atoms, and gives an `effective' density $\tilde{\rho}_s < \rho_s$. According to (\ref{eq:theta}), both effects lead to a lower gas-side contact angle. The solid line in Fig.~\ref{fig:epsgas} shows the predicted contact angle by this expression, assuming an average interaction strength: $\tilde{\epsilon}_{sl} = (\epsilon_{sl}+\epsilon_{lg} )/2$.
Here the effective density is estimated by $\tilde{\rho}_s = (1-\frac{\rho_g^{sl}}{\rho_g^{sat}})\rho_s + \rho_g$, as a phenomenological description for the screening of the solid as the adsorbate layer density grows. 
Note that the influence of the vapour phase is neglected, as was the case in the model without the presence of a gas adsorbate.
The model quantitatively explains the observations in MD, in particular, the saturation of the contact angle occurs exactly when a complete layer of gas adsorbate is formed.
It therefore provides a very natural explanation for the observed universal contact angles in experiments.~\cite{Ishida00,Yang03,Zhang06ca,Limbeek11}

\begin{figure}[t]
\begin{center}
  \includegraphics[width=80mm]{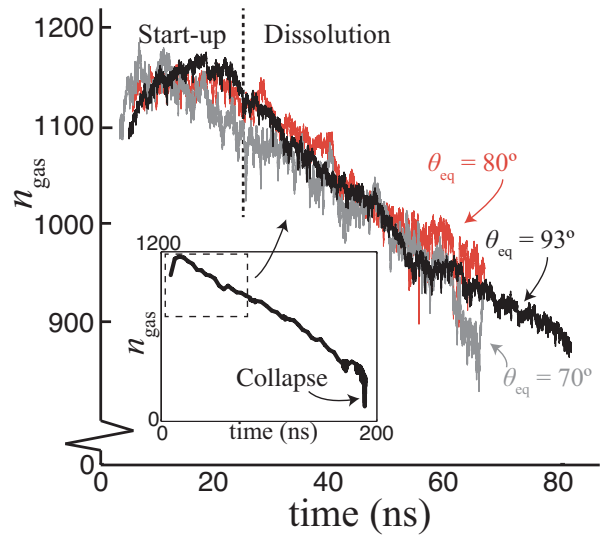}\hspace{3.5cm}
  \caption{\label{fig:time_evol}Number of gas molecules  $n_\textrm{gas}$ inside the nanobubble 
as function of  time for nanobubbles on different substrates with different equilibrium contact angles. Initially, the fluid is supersaturated and a bubble quickly forms within a few~ns. Shortly after the bubble has formed, the liquid is still supersaturated causing gas to enter the bubble (``Start-up"). After about 20~ns the gas in the liquid achieves the equilibrium concentration, and the nanobubbles start to dissolve (``Dissolution"). The dissolution rate of the nanobubbles is independent of the contact angle. The inset shows the full dissolution of the $\theta_{eq} = 93^\circ$ bubble after 0.2~$\mu$s.}
\end{center}
\end{figure}

\paragraph{Stability.} 
Another aspect of nanobubbles that can now be studied is their stability. Are Lennard-Jones nanobubbles stable? After formation of the nanobubble, we use a shape tracker to follow the dynamics of the nanobubble.
The shape tracker locates a nanobubble by performing a circular fit through the liquid-vapour interface of the curved bubble wall.
Different quantities can then be computed, such as the radius of curvature, the contact angle, the amount of gas inside a nanobubble, and the angular dependence of gas flux through the bubble wall. A good indicator for nanobubble stability is the gas content inside the bubble: when the amount of gas remains constant the bubble is considered stable.
The gas contents of nanobubbles on different substrates as a function of time are plotted in Fig.~\ref{fig:time_evol}a. 
We see that none of the nanobubbles are stable; they dissolve on a timescale ($\mu$s) much shorter than that observed in experiments (days), see also the inset in Fig.~\ref{fig:time_evol}.
However, this fast decay is in agreement with simple macroscopic diffusion calculations.
Furthermore, we find that the contact angle of the nanobubbles does not change significantly throughout the dissolution process.

\begin{figure}[t]
\begin{center}
  \includegraphics[width=74mm]{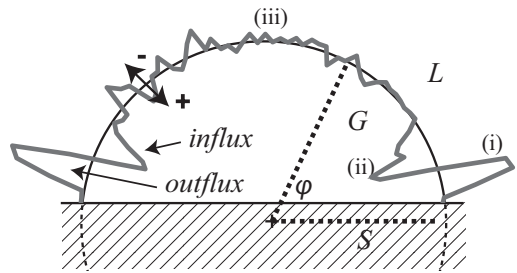}\hspace{3.5cm}
  \caption{\label{fig:localflux}Local flux of gas through the liquid-vapour interface of a nanobubble attached to a substrate. The gray line indicates the time-averaged local gas flow direction and strength. The gas flow is directed outwards everywhere (iii), except for a small region near the contact line [(i) and (ii)] where in a very small region a strong in- and outflux are observed,
 indicating that there exists a recirculation current. The net effect of this recirculation current is found to be of the same order as the diffusive outflux.}
\end{center}
\end{figure}

Although the nanobubbles are not stable, some interesting events occur near the contact line.
When studying the time-averaged local flux as a function of angular position $\phi$ (Fig.~\ref{fig:localflux}b) we see that the flux is highest near the contact line,  indicating that the substrate plays an important role in nanobubble stability.
This strong localized flux near the contact line is heavily influenced by the presence and strength of the gas-enrichment layer, which is a plane at the solid-liquid interface in which gas atoms can move relatively easily due to a liquid depletion layer that exists at the same position. The influx indicates that there may exist a condition where a dynamic equilibrium is achieved, i.e. the diffusive outflux is balanced by the influx at the contact line, explaining why in nonequilibrium surface nanobubbles can be dynamically stable~\cite{Brenner} (and bulk nanobubbles cannot).
A coarse exploration of the parameters $\epsilon_{sl}$ and $\epsilon_{sg}$ has been performed in this study, but 
under the present equilibrium conditions 
a stable nanobubble was not achieved.

Of course, there are many more parameters that need to be explored, such as the initial radius of the bubble: it is possible that nanobubbles below a certain critical size are unstable.
Also, Lennard-Jones fluids might not contain the necessary properties to form stable nanobubbles, such as electrostatic effects.
Most importantly, for the dynamic equilibrium theory to be true, some driving force must exist to sustain the circulation of gas.
This means that the equilibrium simulations in this study need to be adapted to contain such a driving force.
Such non-equilibrium effects include the presence of a thermal gradient (which are likely to be present in experimental setups as well) or the formation of gas at the substrate (which has been studied using electrolysis~\cite{Zhang06,Hui09,Yang09}).

\paragraph{Outlook.} 
In conclusion, we have generated and analysed the formation and stability of surface nanobubbles in simple fluids.
We found that in heavily gas-supersaturated liquids nanobubbles nucleate spontaneously which can then migrate towards the surface. 
In experiments, when water is deposited on the substrate, it is possible that some gas becomes trapped near the solid-liquid interface leading to the required supersaturation.
Other formation mechanisms that cannot be accessed by MD can however not be excluded, further work is required on this question.
The universal contact angles that surface nanobubbles exhibit in experiments can be explained by a dense layer of gas at the solid-liquid interface, which has been shown to exist for real liquids, that effectively alters the substrate chemistry.
Although the Lennard-Jones nanobubbles are unstable, some interesting local gas flows are present near the contact line.
These gas flows are caused by the symmetry breaking due to the solid substrate, and hint towards a dynamic equilibrium condition where the diffusive outflux is compensated by an influx near the contact line.
Since an energy input is required to sustain a circulatory gas flow suggested in the dynamic equilibrium theory by Brenner and Lohse~\cite{Brenner}, it is likely that stable nanobubbles can only occur in non-equilibrium systems.
Simulations of non-equilibrium systems, and of systems containing realistic fluids must be performed to address the question regarding the long lifetime of surface nanobubbles.

\acknowledgements
This work was sponsored by the the NCF (Netherlands National Computing Facilities Foundation) for the use of supercomputer facilities and FOM, both with financial support from the NWO (Netherlands Organisation for Scientific Research).


\begin{thebibliography}{10}

\bibitem{Hampton}
M.~A. Hampton and A.~V. Nguyen, Adv. Colloid Interface Sci. {\bf 154},  30
  (2010).

\bibitem{Seddonrev11}
J.~R.~T. Seddon and D. Lohse, J. Phys.: Condens. Matter {\bf 23},    (2011).

\bibitem{Craig11}
V.~S.~J. Craig, Soft Matter {\bf 7},  40  (2011).

\bibitem{Parker94}
J. Parker, P. Claesson, and P. Attard, {J. Phys. Chem.} {\bf {98}},  8468
  ({1994}).

\bibitem{Bussi07}
G. Bussi, D. Donadio, and M. Parrinello, J. Chem. Phys. {\bf 126},  014101
  (2007).

\bibitem{Note1}
See supplementary material.

\bibitem{Seddon11}
J.~R.~T. Seddon {\it et~al.}, Phys. Rev. Lett. {\bf 106},  056101  (2011).

\bibitem{Zhang04}
X.~H. Zhang {\it et~al.}, Langmuir {\bf 20},  3813  (2004).

\bibitem{Zhang07}
X.~H. Zhang {\it et~al.}, Langmuir {\bf 23},  1778  (2007).

\bibitem{Ishida00}
N. Ishida, T. Inoue, M. Miyahara, and K. Higashitani, Langmuir {\bf 16},  6377
  (2000).

\bibitem{Yang03}
J. Yang, J. Duan, D. Fornasiero, and J. Ralston, The Journal of Physical
  Chemistry B {\bf 107},  6139  (2003).

\bibitem{Zhang06ca}
X.~H. Zhang, N. Maeda, and V.~S.~J. Craig, Langmuir {\bf 22},  5025  (2006).

\bibitem{Weijs10}
J.~H. Weijs {\it et~al.}, {Phys. Fluids} {\bf {23}},  {022001}  ({2010}).

\bibitem{Note2}
Similarly as done in ref.~\cite {Weijs10} for nanodroplets.

\bibitem{GROMOSff}
C.~Oostenbrink, A. Villa, A. E. Mark, and W.F. van Gunsteren, J. Comp. Chem. {\bf 25}, 1656 (2004).

\bibitem{Note3}
A realistic value for the energy between gaseous argon and individual carbon
  atoms in graphene is: $\epsilon _{Ar\protect \tmspace +\thickmuskip
  {.2777em}-\protect \tmspace +\thickmuskip {.2777em}C} = 0.2$~$k_bT$ (at 300~K). For the
  energies between the oxygens of two water molecules: $\epsilon _{O_l\protect
  \tmspace +\thickmuskip {.2777em}-\protect \tmspace +\thickmuskip
  {.2777em}O_l} = 0.26$~$k_bT$.\cite {GROMOSff}.

\bibitem{Dammer}
S.~M. Dammer and D. Lohse, Phys. Rev. Lett. {\bf 96},  206101  (2006).

\bibitem{Bratko08}
D. Bratko and A. Luzar, Langmuir {\bf 24},  1247  (2008).

\bibitem{Sendner09}
C. Sendner, D. Horinek, L. Bocquet, and R.~R. Netz, Langmuir {\bf 25},  10768
  (2009).

\bibitem{Rowlinson/Widom}
J.~S. Rowlinson and B. Widom, {\em Molecular Theory of Capillarity} (Dover
  publications, ADDRESS, 1982).

\bibitem{BauerDietrich}
C. Bauer and S. Dietrich, Eur. Phys. J. B {\bf 10},  767  (1999).

\bibitem{Snoeijer}
J.~H. Snoeijer and B. Andreotti, {Phys. Fluids} {\bf {20}},  {057101}
  ({2008}).

\bibitem{Limbeek11}
M.~A.~J. van Limbeek and J.~R.~T. Seddon, Langmuir {\bf 27},  8694  (2011).

\bibitem{Brenner}
M.~P. Brenner and D. Lohse, Phys. Rev. Lett. {\bf 101},  214505  (2008).

\bibitem{Zhang06}
L. Zhang {\it et~al.}, Langmuir {\bf 22},  8109  (2006).

\bibitem{Hui09}
F. Hui {\it et~al.}, Electrochemistry Communications {\bf 11},  639   (2009).

\bibitem{Yang09}
S. Yang {\it et~al.}, Langmuir {\bf 25},  1466  (2009).


\end{thebibliography}
\end{document}